\documentclass{article} 
\usepackage{emulateapj} 
\usepackage[]{epsfig} 
\hoffset -1.5truecm \lefthead{Menci et al.} 
\righthead{} 
\def\newline{\hfil\break} 
\begin{document} 
\title{X-Ray Evolution of Active Galactic Nuclei \\ 
and Hierarchical Galaxy Formation} 
\author{N. MENCI\altaffilmark{1}, F. FIORE\altaffilmark{1}, 
G.C. PEROLA\altaffilmark{1,2}, A. CAVALIERE\altaffilmark{3}}
\altaffiltext{1}{INAF - Osservatorio Astronomico di Roma,
via di Frascati 33, 00040 Monteporzio, Italy}
\altaffiltext{2}{Dipartimento di Fisica, Universitá di Roma Tre, 
Via della Vasca Navale 84, I-00146 Roma, Italy}
\altaffiltext{3}{Dipartimento Fisica, II Universit\`a di Roma,
via Ricerca Scientifica 1, 00133 Roma, Italy}
 
\begin{abstract} 
We have incorporated the description of the X-ray properties of Active
Galactic Nuclei (AGNs) into a semi-analytic model of galaxy
formation, adopting physically motivated scaling laws for
accretion triggered by galaxy encounters. 
Our model reproduces the level of the cosmic X-ray
background at 30 keV; we predict that the largest contribution
(around $2/3$) comes from sources with intermediate X-ray
luminosity $10^{43.5}< L_X/{\rm erg\,s^{-1}} <10^{44.5}$, with 50 \% 
of the total specific intensity produced at $z<2$. The 
predicted number density of luminous 
X-ray AGNs ($L_X>10^{44.5}$ erg/s in the 2-10 keV band) peaks 
at $z\approx 2$ with a decline of around 3 dex to $z=0$;  
for the low luminosity sources
($10^{43}<L_X/{\rm erg\,s^{-1}}<10^{44}$) it has a broader and less 
pronounced maximum around $z\approx 1.5$. 
The comparison with the data shows a generally good 
agreement. The model predictions slightly exceed the observed number of 
low-luminosity AGNs at $z\sim 1.5$, with the discrepancy progressively extending 
to intermediate-luminosity objects at higher redshifts; we 
discuss possible origins for the
mismatch. Finally, we predict the source counts and the flux
distribution  at different redshifts in the hard (20-100 keV) X-ray band for the sources
contributing to the X- ray background. 
\end{abstract} 

\keywords{galaxies: active -- galaxies: formation -- X-rays: diffuse
background -- X-rays: galaxies -- galaxies: evolution -- galaxies:
interactions}

\section {Introduction} 

The X-ray emission of Active Galactic Nuclei (AGNs) provides a unique
probe into the history of the accretion on supermassive black holes
(BHs) thought to power the AGNs. In fact, the X-rays probe the
accretion down to BH masses as small as $m_{BH}\sim
10^7\,M_{\odot}$ and to bolometric luminosities as low as $L\sim
10^{43}$ erg/s.  At redshifts $z\gtrsim 0.5$, these regimes of
accretion are hardly accessible to optical observations, but may
provide both a significant fraction of the total accretion power in
the Universe and a considerable contribution to the observed Cosmic
X--ray Background (CXB).

At higher $z$, constraints on the properties of the AGNs, and hence on
the history of accretion, have been derived by extrapolating local
properties like the optical or X-ray luminosity functions, or the
X-ray spectral shape of nearby Seyfert galaxies. Based on these
assumptions, several authors (see Madau, Ghisellini \& Fabian 1994;
Comastri et al. 1995, 2001; Wilman \& Fabian 1999) have constructed
AGN synthesis models to account for the CXB; these authors argued that
most of the accretion power producing the CXB at energies below 30 keV
is intrinsically absorbed, as initially suggested by Setti \& Woltjer
(1989).

The drawbacks of such an approach are constituted by the degeneracies
between different possible extrapolations and hence between synthesis
models that still reproduce the observed integrated AGN
properties. Furthermore, little direct insight can be gained on the
physical links between the evolution of AGNs and that of their host
galaxies.  To now, the incorporation of the X-ray properties of AGNs
into a coherent picture of galaxy evolution is still lacking.

On the other hand, the attempt by Kauffmann \& Haehnelt (2000)
to connect the BH accretion to the evolution of galaxies in the
hierarchical scenario was based on a semi-analytic model (SAM) of
galaxy formation of the kind introduced by Kauffmann et al. (1993) and
Cole et al. (1994) and progressively developed by several authors
(Somerville \& Primack 1999; Wu, Fabian \& Nulsen 2000; Cole et
al. 2000, Menci et al. 2002). Such a model assumed phenomenological
and tunable scaling laws to connect the gas accreted onto the BH with
the cold gas fraction available in the galaxies left over by the star
formation. However, the model has been applied only to the bright
quasar (QSO hereafter) population, yet it has failed to reproduce the
fast drop (by a factor $\sim 10^{-2}$) of their density observed from
$z\approx 2$ to the present. A later paper by Nulsen \& Fabian (2000)
assumed that the BHs are specifically fueled by the cooling flows
associated with the hot gas during the process of hierarchical galaxy
evolution, the latter being described through a SAM. However, such a
model fails to account for the observed relationship between the black
hole and the spheroid masses (see Richstone et al.  1998); furthermore, it  
does not properly describe the statistical distribution of QSO properties in
that it over-predicts their low-$z$ luminosity functions and
under-predicts QSOs at high $z$. Other analytic (see, e.g., 
Haiman \& Loeb 1998; Wyithe \& Loeb 2002; Hatziminaoglou et al. 2003), 
semi-analytic (Volonteri, Hardt \& Madau 2003) 
or N-body (Kauffmann \& Haehnelt 2002) works are based on major merging events 
of dark matter haloes at high $z$ as the only triggers for BH accretion, 
which is related to the DM mass of the halo either adopting the locally 
observed relations or through phenomenological scaling laws;  
thus such models are suited to follow the QSO evolution at high-intermediate $z$, 
but do not match the observed steep decline of the QSO density 
at redshifts $z\lesssim 1$, a range of cosmic time of 
key interest for the activity of AGNs observed in X-rays. 

Recently, Menci et al. (2003) developed a more physical model to
connect the BH accretion to the galaxy evolution in the hierarchical
scenario. The accretion is triggered by galaxy encounters,  
not necessarely leading to bound merging, in common
host structures like clusters and especially groups; these events destabilize
part of the galactic cold gas and hence feed the central BH, following
the physical modelling developed by Cavaliere \& Vittorini (2000,
CV00). The amount of the cold gas available, the interaction rates,
and the properties of the host galaxies are derived through the SAM
developed by Menci et al. (2002).

As a result, at high $z$ the protogalaxies grow rapidly by
hierarchical merging; meanwhile, much fresh gas is imported and also
destabilized, so the BHs are fueled at their full Eddington rates. At
lower $z$, the dominant dynamical events are galaxy encounters
in hierarchically growing groups; now refueling peters out, as the
residual gas is exhausted while the destabilizing encounters
dwindle. With no parameter tuning other than needed for star
formation in canonical SAMs, the model naturally produces in the
bright QSO population a rise for $z>3$, and for $z\lesssim 2.5$ a drop
as steep as observed. In addition, the results closely reproduce the
observed luminosity functions of the optically selected QSOs, their
space density at different magnitudes from $z\approx 5$ to $z\approx
0$, and also the local $m_{BH}-\sigma$ relation.

Encouraged by the successes of the model, here we extend it to lower
accretions and lower BH masses, in order to study the X-ray properties
of the AGNs, and to predict their contribution to the X-ray
background. We avoid the introduction in our model of dust and gas
obscuration, and therefore conservatively we shall compare our
predicted CXB with the observations at energies $E\geq 30$ keV, the
spectral region not affected by photoelectric absorption. At lower
energies we shall compare our predicted number and luminosity
densities of AGNs with data already corrected for dust and gas
absorption.
 
The paper is organized as follows. In Sect. 2 we recall the basic
points of the galaxy evolutionary model we base on. The specific
modelling that relates the BH accretion to the galaxy encounters is
recalled in Sect. 3, where we also describe how we obtain the X-ray
luminosities. The results concerning the X-ray properties of AGNs at
low and high $z$ are compared with observations in Sect. 4.  The final
Sect. 5 is devoted to discussion and conclusions.

\section{The Model}

\subsection{The Semi-analytic Model for Galaxy Evolutiuon}

The properties of the galaxies hosting the AGNs are derived using the
semi-analytic model described in detail in Menci et al. (2003).  Here
we recall its basic points.

Following the procedure usually adopted in SAMs, we consider both the
host dark matter (DM) halos (groups and clusters of galaxies with mass
$M$, virial radius $R$ and circular velocity $V$) containing the
galaxies, and the DM clumps (with circular velocity $v$) associated
with the individual member galaxies. The former grow hierarchically to
larger sizes through repeated merging events (at the rate given, e.g.,
in Lacey \& Cole 1993). On the other hand, the latter may coalesce
either with the central galaxy in the common halo due to dynamical
friction, or with other companion galaxies through binary
aggregations. We Compute the probability for the latter processes to
occur during the hierarchical growth of the hosting structure yields,
at any cosmic time $t$, and the differential distribution $N(v,V,t)$
(per Mpc$^3$) of galaxies with given $v$ in hosts with circular
velocity $V$.  From this we derive the number $N_T(V,t)$ of galaxies
in a host halo (i.e., the membership), and the overall distribution of
galaxy circular velocity $N(v,t)$ irrespective of the host halo. To
each galactic circular velocity $v$, it is associated an average
galaxy tidal radius $r_t(v)$; the average relative velocity $V_r(V)$
of the galaxies in a common DM halo is computed for each circular
velocity $V$ of the host halo.

The properties of the gas and stars contained in the galactic DM
clumps are computed as follows. Starting from an initial gas amount
$m\,\Omega_b/\Omega$ ($m\propto v^3$ being the DM mass of the
galaxies) at the virial temperature of the galactic halos, we compute
the mass $m_c$ of cold baryons residing in regions interior to the
cooling radius.  The disk associated to the cold baryons has radius
$r_d(v)$ and rotation velocity $v_d(v)$ computed after Mo, Mao \&
White (1998).  From such a cold phase, stars are allowed to form with
rate $\dot m_* = (m_c/ t_{d})\,(v/ 200\,{\rm km\,s^{-1}})^{-\alpha_*}$
with the disk dynamical time evaluated as $t_d = r_d/v_d$.

Finally, a mass $\Delta m_h=\beta\,m_*$ is returned from the cool to
the hot gas phase due to the energy fed back by canonical type II
Supernovae associated to $m_*$. The feedback efficiency is taken to be
$\beta= (v/v_h)^{\alpha_h}$; the values adopted for the free
parameters $\alpha_*=-1.5$, $\alpha_h=2$ and $v_h=150$ km/s fit both
the local B-band galaxy LF and the Tully-Fisher relation, as
illustrated by Menci et al. (2002). The model also matches the bright
end of the galaxy B-band luminosity functions up to redshifts
$z\approx 3$ (see Poli et al. 2003) and the resulting global star
formation history is broadly consistent with that observed up to
redshift $z\approx 4$ (Fontana et al. 1999).

At each merging event, the masses of the different baryonic phases are
replenished by those in the merging partner; the further increments
$\Delta m_c$, $\Delta m_*$, $\Delta m_h$ from cooling, star formation
and feedback are recomputed on iterating the procedure described
above.

Thus, for each galactic circular velocity the star formation 
described above is driven by the cooling rate of the hot gas and by
the rate of replenishing the cold gas, which in turn is related to the
progressive growth of the total galactic mass along the galaxy merging
tree. The integrated stellar emission $S_{\lambda}(v,t)$ at the
wavelength $\lambda$ is computed by convolving the star formation rate
with the spectral energy distribution $\phi_{\lambda}$ obtained from
population synthesis models (Bruzual \& Charlot 1993, and subsequent
updates).

All computations are made in a $\Lambda$-CDM cosmology with
$\Omega_0=0.3$, $\Omega_{\lambda}=0.7$, a baryon fraction
$\Omega_b=0.03$, and Hubble constant $h=0.7$ in units of 100 km
s$^{-1}$ Mpc$^{-1}$.

\subsection{The Accretion onto Supermassive BHs}

We follow the model presented in Menci et al. (2003) to derive the
fraction of the galactic cold gas accreted onto the central BH.
Galaxy encounters are expected to destabilize part of such 
gas causing it to loose angular momentum (Mihos \& Hernquist
1996; Barnes \& Hernquist 1998; see also Mihos 1999), and thus
triggering gas inflow. 

The fraction of cold gas destabilized in each interaction event is
computed in eq. A3 of CV00 in terms of the variation $\Delta j$ of the
specific angular momentum $j\approx Gm/v_d$ of the gas, to read
\begin{equation}
f(v,V)\approx {1\over 8}\, \Big|{\Delta j\over j}\Big|= {1\over
8}\Big\langle {m'\over m}\,{r_d\over b}\,{v_d\over V}\Big\rangle\, .
\end{equation}
Here $b=max[r_d,R/N_T^{-1/3}(V)]$ is the impact parameter, $m'$ is the
mass of the partner galaxy in the interaction, and the average runs
over the probability of finding a galaxy with mass $m'$ in the same
halo $V$ where the galaxy $m$ is located.  The prefactor accounts for
the probability (1/2) of inflow rather than outflow related to the
sign of $\Delta j$. 
In addition, the gas funneled inward may end up also 
in a nuclear starburst; the relative amount has been estimated from 
3/1 to 9/1 (see Sanders \& Mirabel 1996; Franceschini, Braito \& Fadda 2002).
Here we assume that $1/4$ of the inflow feeds the central BH, while the 
remaining fraction kindles circumnuclear starbursts, tackled in
detail by Menci et al. (2004).

The rate of nearly grazing encounters for a galactic halo with given
$v$ inside a host halo (group or cluster) with circular velocity $V$,
is given by $\tau_r^{- 1}=n_T(V)\,\Sigma (v,V)\,V_r(V)$; here
$n_T=N_T/( 4 \pi$$R^3/3)$, and the cross section is
$\Sigma(v,V)\approx\pi \langle \,(r_t^2+r_t^{'2})\rangle$ is averaged
over all partners with tidal radius $r'_t$ in the same halo $V$. The
membership $N_T$, and the distributions of $v'$, $r_t'$ and $V_r$ are
computed from the SAM as described in Sect. 2.

The average gas accretion rate triggered by interactions at $z$ is
given by (cf. CV00 eq. 5, and Menci et al. 2003)
\begin{equation}
\dot m_{acc}(v,z)=\Big\langle{f(v,V)\,\,m_c(v)\over \tau_r(v,V)}\Big\rangle ~, 
\end{equation}
where the average is over all host halos with circular velocity $V$.
The mass of the BH hosted in a galaxy with given $v$ at time $t$ is
updated after $m_{BH}(v,t)=(1-\eta)\int_0^t\,\dot m_{acc}(v,t')\,dt'$,
where $\eta\approx 0.1$ (see Yu \& Tremaine 2002) is the mass-energy
conversion efficiency; here we assume in all galaxies initial seed BHs
of mass much smaller than the active supermassive BHs (see
Madau \& Rees 2001).

The bolometric luminosity so produced by the accretion onto a BH
hosted in a galaxy with given $v$ then reads
\begin{equation}
L(v,t)={\eta\,c^2\Delta m_{acc}\over \tau} ~. 
\end{equation}
Here $\tau \approx t_d \sim 5\,10^7\,(t/t_0)$ yrs is the duration of
the accretion episode, i.e., the timescale for the QSO to shine;
$\Delta m_{acc}$ is the gas accreted at the rate given by eq. (2).
The blue luminosity $L_B$ is obtained by applying a bolometric
correction 13 (Elvis et al. 1994), while for the unabsorbed X-ray
luminosity $L_X$ (2-10 keV) we adopt a bolometric correction
$c_{2-10}=100$ following Elvis, Risaliti \& Zamorani (2002); for
simplicity, this is assumed to be constant with $z$. The shape of the
X-ray spectrum $I(E)$ is assumed to be a power law with a slope
$\alpha=0.9$ (see Comastri 2000 and references therein), with an
exponential cutoff at an energy $E_c=300$ keV (see e.g. Perola et
al. 2002 and references therein); in view of the present data
situation we shall keep this as our fiducial shape. The effect of the
scatter in $\alpha$ and $E_c$ will be discussed in Sect. 4.

The evolving LF is derived from $N(v,t)$ by applying the appropriate
Jacobian given by eq. (3) with the appropriate bolometric
correction. The LF will include a factor $\tau
/\langle\tau_r(v)\rangle < 1$ since the luminosities in eq. (3) last
for a time $\tau$ and are rekindled after an average time
$\tau_r$. The result is
\begin{equation} 
N(L,t)=N(v,t)\,{\tau\over \langle\tau_r\rangle}\,\Big|{dv\over dL}\Big|\,.
\end{equation} 

As shown by Menci et al. (2003), the model produces a drop of the
bright ($M_B<- 24$) QSOs as steep as observed. This results from the
combined decrease with time of both the interaction rate $\tau_r^{-1}$
and the accreted fraction $f$ (see Fig. 1 in Menci et al. 2003). These
in turn are caused by the decrease of the encounter probability due to
the decrease of the number density of galaxies inside the host
groups/clusters, and to the simultaneous increase of galaxy relative
velocities $V_r$. For massive galaxies, the above effects also combine
with the decrease of the available amount of cold gas, due to its
rapid conversion into stars in the early phases of star formation.

The above effects concur to produce number densities and luminosity
functions of QSOs in excellent agreement with the observations in the
full range $0<z<6$, as shown in Menci et al. (2003). But comparison
with optical data allows to probe the model only in the medium-high
accretion regimes, corresponding to optical magnitudes brighter than
$M_B\approx -24$. Here we explore the predictions of our model down to
accretion rates lower by a factor $\sim 10^{-1}$, which at present can
only be probed by X-rays.

\section{The Overall Picture}

Before presenting in detail our results and comparing
them with observations, we give an overview of the picture concerning
the BH accretion in evolving galaxies as it emerges from our model. In
Fig. 1 we show the volume emissivity from the AGNs which at redshift
$z$ have luminosity $L_X$. Since our model links the evolution of AGNs
to that of the hierarchically growing galaxies, we can gain some
insight also on the hosts of the AGNs contributing to the accretions
history of the Universe. So we also show in the figure the
evolutionary tracks of AGNs hosted in galaxies with different masses.

At redshifts from 6 to 3 a rapid increase of the accretion is
sustained by the continuous replenishing of cold gas due to the frequent
galaxy merging events and to the high rate of encounters which
destabilize it. The tradeoff bewteen high luminosity and large number
of sources determines a peak in the global emissivity at intermediate
AGN luminosities $L_X\sim 10^{44}$ erg/s, which are hosted typically
in galaxies with DM mass $5\,10^{10}-3\,10^{11}\,M_{\odot}$.

At later $z$ the era of galaxy formation ends, the merging rate of
galaxies drops, and the cold gas replenishing dwindles. The encounter
rate of galaxies also decreases, and the galaxies enter a era of
nearly passive evolution. The AGNs in the most massive galaxies (with
DM masses $M>10^{12}\,M_{\odot}$) show the most dramatic decrease in
luminosity (see the leftmost track in Fig. 1), since their host
galaxies have already converted most of their gas into stars at higher
$z$. The decrease at $z\lesssim 1.8$ of the luminosities of the AGN
population produces the decrease in the total emissivity shown in
Fig. 1.

The location of the peak in both redshift and luminosity, as well as
the specific evolution with $z$ of the luminosity of AGNs hosted in
galaxies with different masses, strongly affects the relative
contribution of different classes of AGNs to the CXB and the redshift
distribution of their number or luminosity density, as we show in
detail below.

\begin{center}
\vspace{-0.9cm} 
\scalebox{0.33}[0.33]{\rotatebox{270}{\includegraphics{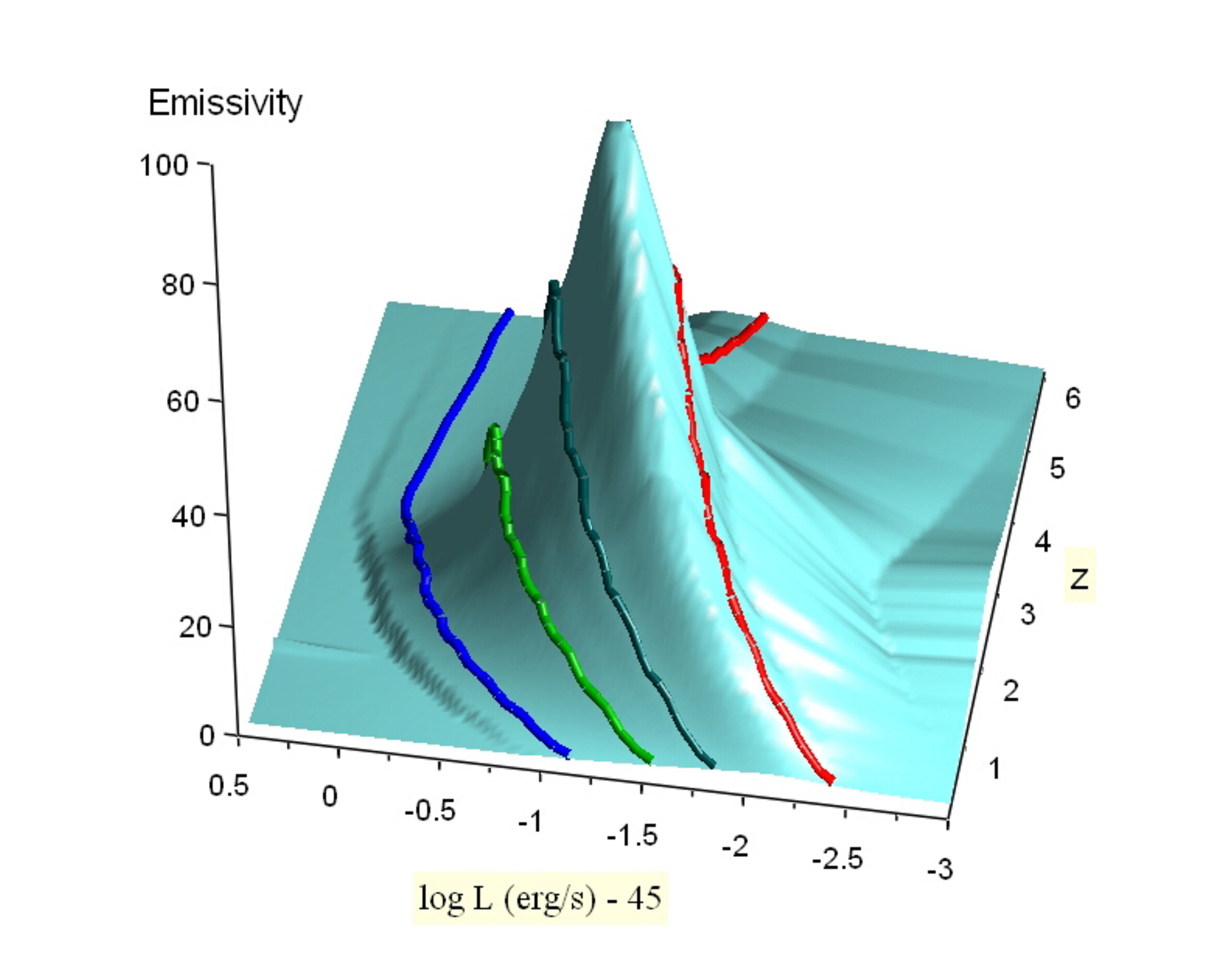}}}
\end{center}
{\footnotesize
\vspace{-0.3cm } Fig. 1. - The overall X-ray emissivity contributed
by AGNs with given X-ray luminosity $L_X$ and redshift $z$. We also show the 
evolutionary tracks of the luminisity of AGNs hosted in galaxies with different DM
masses corresponding (from left to right) to $10^{13}\,{\rm M}_{\odot}$, $10^{12}\,{\rm M}_{\odot}$, 
$2.5\,10^{11}\,{\rm M}_{\odot}$, $5\,10^{10}\,{\rm M}_{\odot}$.  
\vspace{0.cm}}

\section{Results}

We compute the contribution to the CXB from the LFs in eq. (4) as
follows:
\begin{equation}
{d^2 F(E)\over dE\,d \omega}= \int\,dV\, \int_{L_1}^{L_2} dL_X\,
N(L_X,z)\,{I[L_X,E(1+z)]\over 4\pi\,d_L^2}\,.
\end{equation}
Here $dV$ is the cosmic volume element per unit solid angle $\omega$,
$d_L$ is the luminosity distance, and $I(L,E)$ is the AGN
spectrum. This is normalized as to yield $L_X=L/c_{2-10}$ when
integrated in the rest-frame energy range 2-10 keV.  We consider the
contribution to the CXB from AGNs in specific luminosity ranges
$\Delta L=[L_1,L_2]$. The total CXB is obtained upon integrating over
the full range of luminosities $10^{42}<L_X/{\rm erg/s}<5\,10^{46}$
spanned by our SAM.

Since the model does not include obscuration, we shall directly
compare the results from eq. 5 with the hard CXB observed at $E\geq
30$ keV where photoelectric absorption does not affect the observed
fluxes.  Nonetheless, we must be aware of the fact that, even at this
energy, the value of the observed background is appreciabely affected
by the presence of sources in a Compton thick phase of obscuration.

In Fig. 2 we show the hard CXB at $E_0=30$ keV obtained from eq. (5)
by integrating the volume out to running redshift $z$ .  We show both
the global value, and the fraction contributed by AGNs in three
classes of luminosity.  The predicted background with the chosen
parameters for the spectrum ($\alpha=0.9$, $E_c=300$ keV,
$c_{2-10}=100$, see Sect. 2.2) exceeds the value measured by HEAO1-A2
by no more than 50 \%.

At the present state of our observational knowledge, the substantial
agreement is very encouraging, especially if the following points are taken
into account: a) the available evidence (at $E<10$ keV, 
Lumb et al. 2002;  Vecchi et al. 1999) that the CXB normalization from the HEAO1-A2 experiment
may be underestimated by as much as 30 \%; b) the bolometric correction
need take on the fixed value we adopted for all values of $L$
and $L/L_{edd}$; c) the incidence of a Compton thick phase along the
active phase of a galactic nucleus, as a function of $L$ and $z$, is
not known, except that locally it may amount as much as 50 \%
(Risaliti, Maiolino \& Salvati 1999) of the so-called type 2 AGNs,
namely those with a substantial obscuration both in the optical as
well as in the X-ray band.  The essential features of our predictions
are shown in figs. 2 and 3.

\begin{center}
\vspace{0cm} 
\scalebox{0.45}[0.45]{\rotatebox{0}{\includegraphics{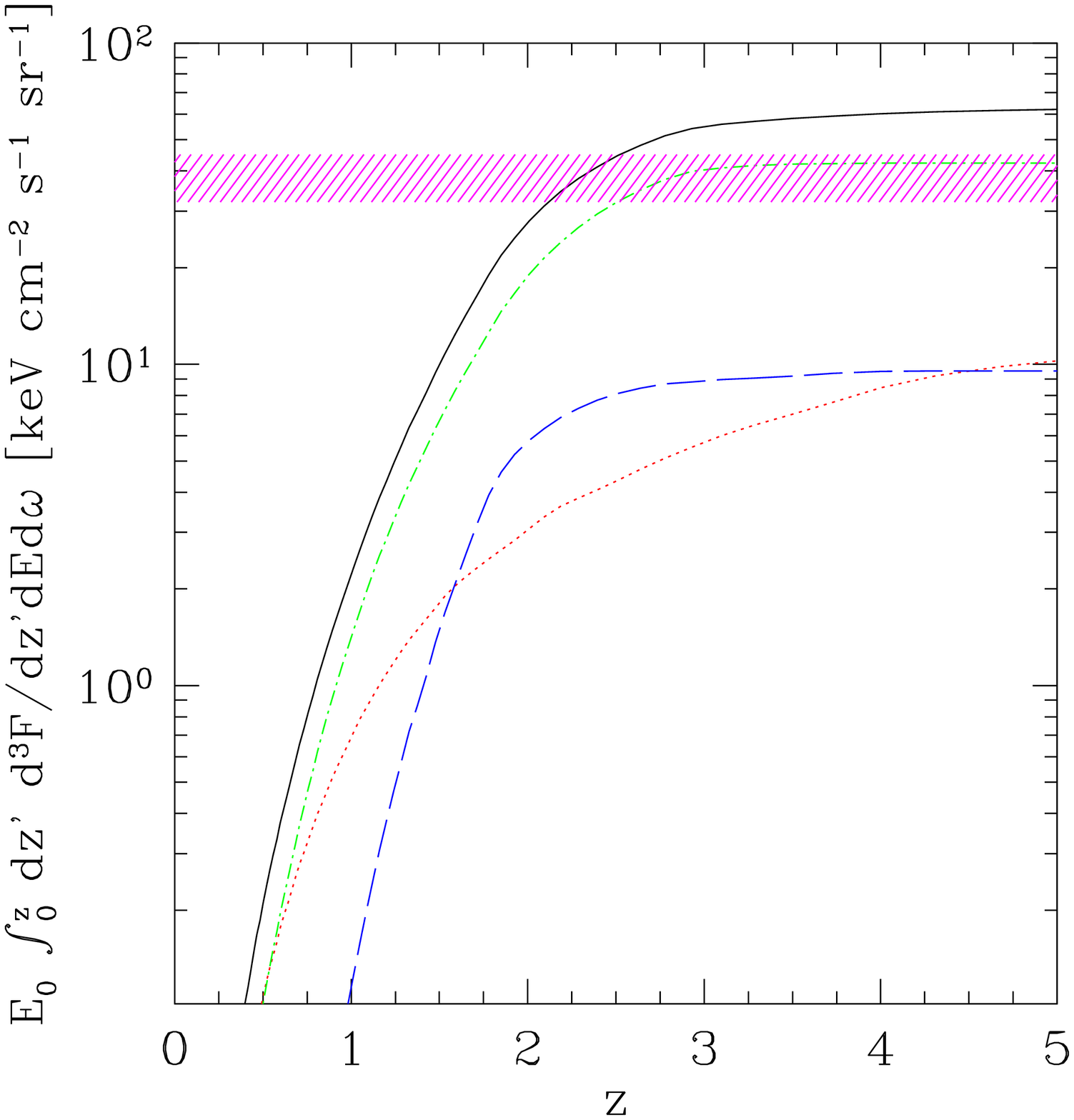}}}
\end{center}
{\footnotesize
\vspace{0cm } Fig. 2. - The cumulative contruibution (multiplied by
the energy $E_0$) to the predicted CXB at $E_0=30$ keV, yielded by
sources at progressively larger redshifts.  The solid line shows the
total CXB produced by sources with all luminosities. The other lines
show the contributions of AGNs with luminosities $L_x$ (in units of
erg $s^{-1}$ in the band 2-10 keV) in the ranges $42<log L_X<43.5$
(dotted), $43.5<log L_X<44.5$ (dot dashed), and $44.5<log L_X$
(long-dashed). The shaded strip is the value 43 keV cm$^{-2}$ s$^{-1}$
sr$^{-1}$ measured by HEAO1-A2 (Gruber et al. 1999).
\vspace{0.3cm}}

Fig. 2 shows that in our model the CXB is mainly contributed by AGNs
with intermediate luminosities $L_X=10^{43.5}-10^{44.5}$ erg/s, which
provide $\approx 50\%$ of the total value. High luminosity
$(L_X>10^{44.5}$ erg/s) and low luminosity ($L_X<10^{43}$ erg/s)
sources contribute a fraction $\sim 25 \%$ each.  The population with
intermediate luminosities strikes the best tradeoff between larger
luminosity and smaller number of sources.  Thus, in this picture high
luminosity highly absorbed objects (the so-called type 2 QSOs) would
not give a dominant contribution to the hard CXB. In fact, although
recent XMM and Chandra surveys are providing a sizeable number of QSO2
(Barger et al. 2002, Fiore et al. 2003, Hasinger 2003), these are
likely to constitute a relatively minor fraction of sources down to
the fluxes where the bulk of the hard CXB is resolved into sources.

\begin{center}
\vspace{0cm} 
\scalebox{0.43}[0.43]{\rotatebox{0}{\includegraphics{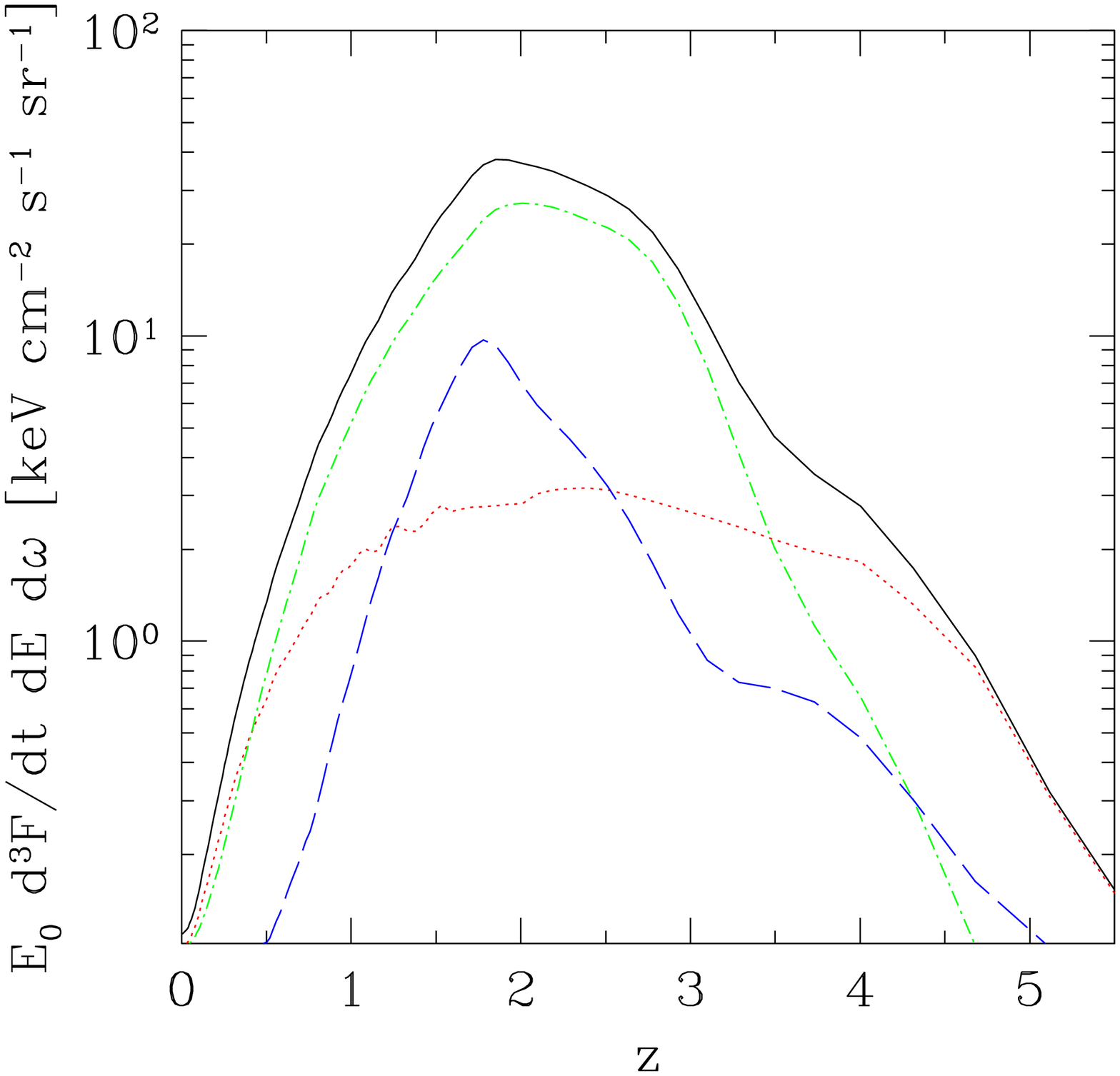}}}
\end{center}
{\footnotesize
\vspace{0.cm } 
Fig. 3. - It is shown the differential contruibution to the predicted 
CXB at $E_0=30$ keV, for different ranges of luminosity of the contributing sources.
Symbols are as in Fig. 2
\vspace{0.3cm}}

\begin{center}
\vspace{0cm} 
\scalebox{0.43}[0.43]{\rotatebox{0}{\includegraphics{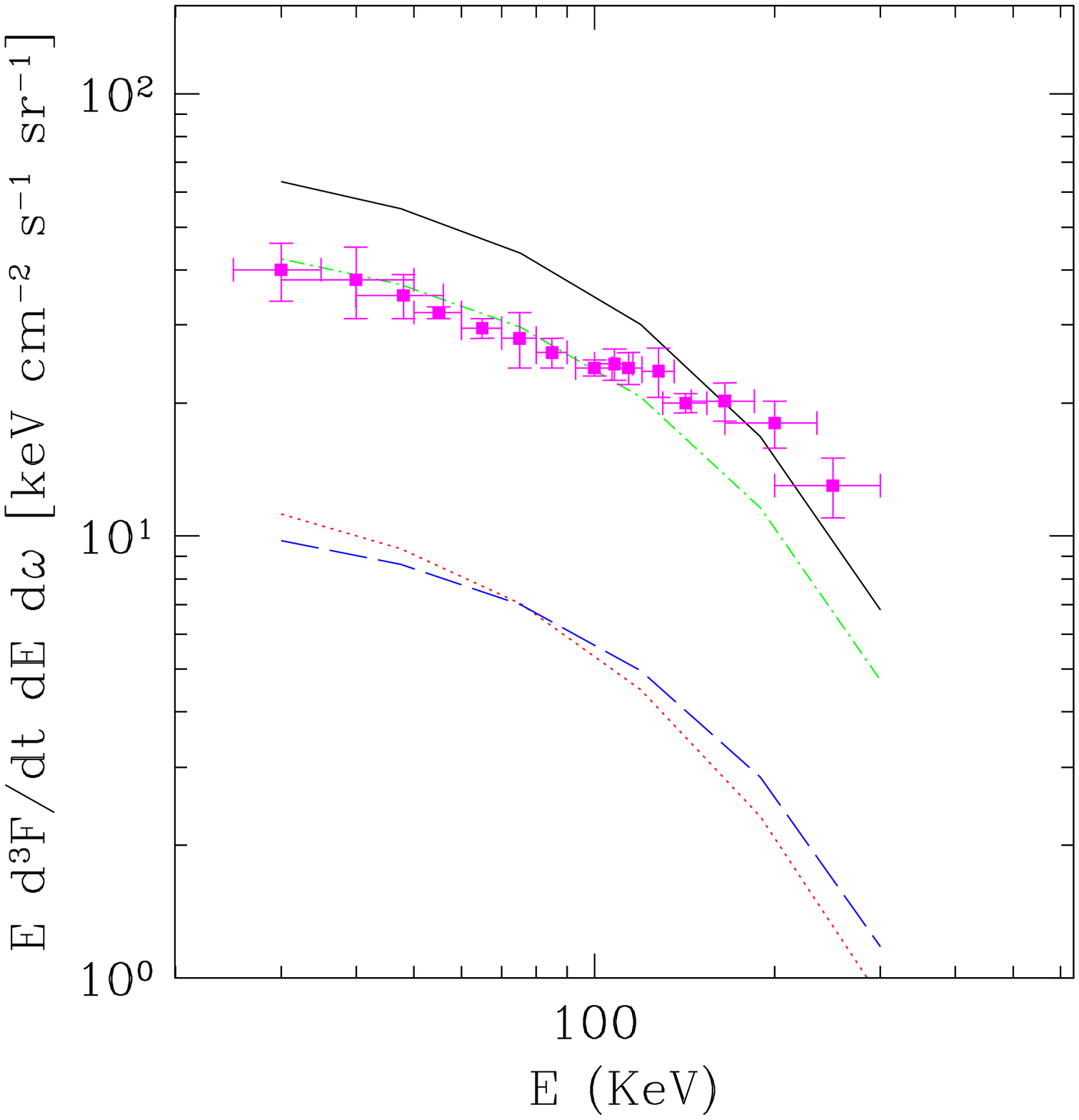}}}
\end{center}
{\footnotesize
\vspace{0.cm } Fig. 4. - The energy spectrum of the predicted CXB
for $E\geq 30$ keV, for different ranges of luminosity of the
contributing sources.  Symbols are as in Fig. 1. Data are from Gruber
et al. (1999).
\vspace{0.3cm}}

Note also that, while the contribution of the high luminosity sources
saturates already at $z\approx 2$, that from low luminosity ones
continues to rise up to $z\approx 3-4$. This behavious can be regarded
as a natural outcome in the framework of a hierarchical scenario,
where at earlier times increasingly largeer numbers of small galaxies
are present, which later merge to form larger systems.

Further insight into the redshift distribution of the contribution to
the CXB at 30 keV is provided in Fig. 3, which shows the
$z$-derivative of the CXB in Fig. 2. The contribution to the CXB from
all sources peaks at $z\approx 2$, and so does the contribution of the
intermediate luminosity sources. The contribution of high luminosity
AGNs, instead, is more sharply peaked at the slightly lower redshift,
$z\approx 1.7$, while the contribution from the lowest luminosity
sources shows a broad $z$-distribution.

The CXB spectrum above 20 keV, computed with the fixed values of
$\alpha$ and $E_c$ given above, is compared with the HEAO1-A2 data in
Fig. 4. Apart from the normalization mismatch commented previously,
the predicted shape is slightly softer than observed. We note however
that, in a more detailed modelling, the adoption of an appropriate
scatter in $\alpha$ and $E_c$ is likely to yield a somewhat harder
shape.

\begin{center}
\vspace{0.cm} 
\scalebox{0.45}[0.45]{\rotatebox{0}{\includegraphics{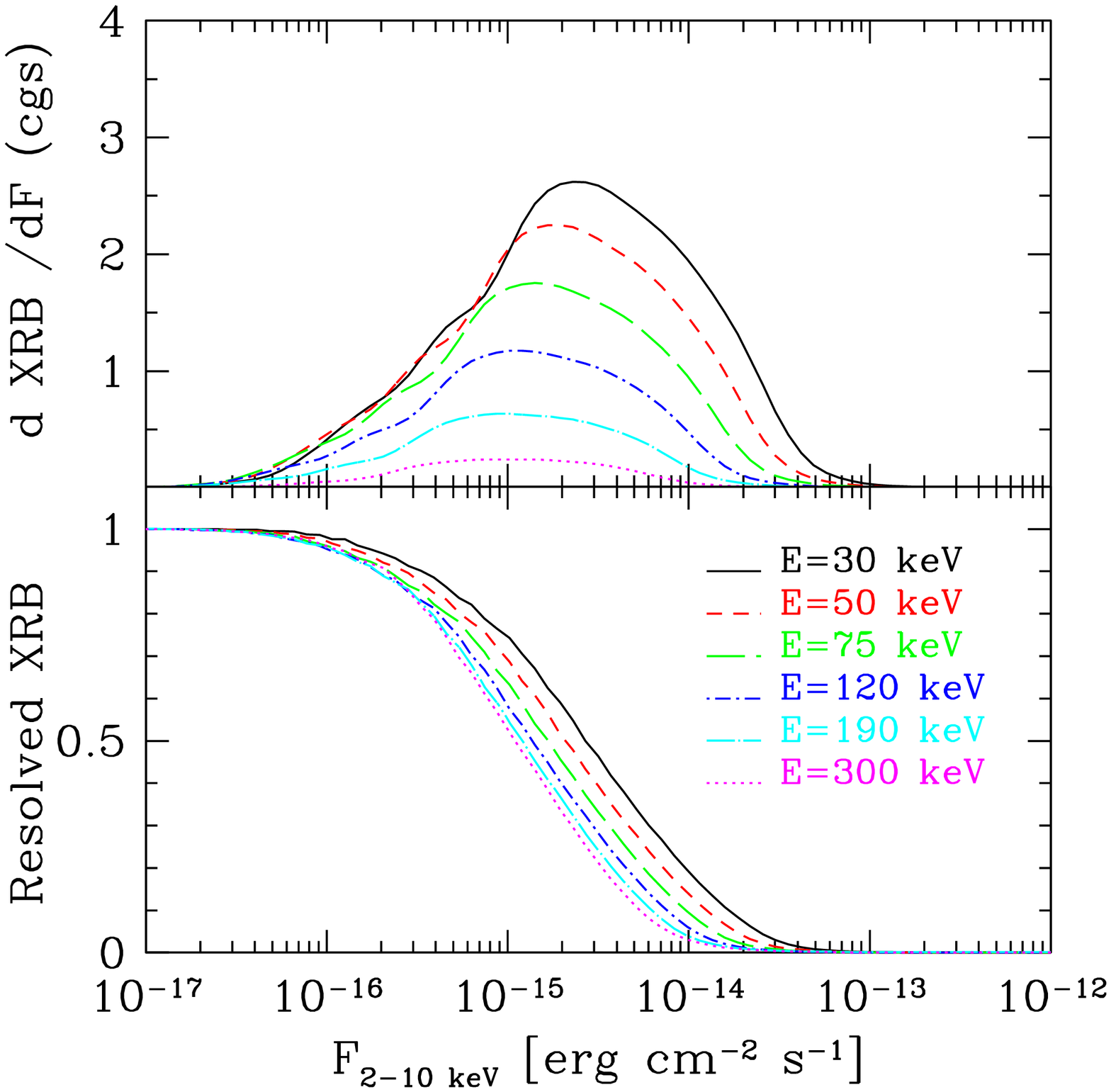}}}
\end{center}
{\footnotesize
\vspace{-0.1cm } Fig. 5. - Top panel. The flux distribution of AGNs
contributing to the CXB at different energies (see labels in the
bottom panel). The flux corresponds to the intrinsic emission
(corrected for obscuration) in the currently accessible energy band
2-10 keV.  Bottom panel. The fraction of the CXB contributed by
sources with fluxes larger than the considered value.
\vspace{0.3cm}}

Our model can provide guidelines for future observations.  Fig. 5
shows the differential (upper panel) and integral (lower panel)
contribution to the CXB at various energies (from 30 to 300 keV)
provided by sources with different 2-10 keV flux (corrected for
obscuration).  As a function instead of their 20-100 keV flux, the
panels in Fig. 6 show the differential and the integral contributions
to the CXB at 30 keV in various bins of redshifts, from $z=0-1.5$ to
$z>3.5$. This figure illustrates the predictions that the bulk of the
CXB should be provided by sources with fluxes around $F_{20-100 {\rm
keV}}\approx 10^{-14}$ ergs cm$^{-2}$ s$^{-1}$, placed at redshifts
$z\approx 1.5 -2.5$.  To observationally access this flux levels, high
energy focussing optics is required, a tchnological step forward which
has become feasible and is included in current studies for further
missions, like, e.g., Constellation X
\footnote{http://constellation.gsfc.nasa.gov/} and NeXT (Tawara et al. 2003).

\begin{center}
\vspace{0cm} 
\scalebox{0.45}[0.45]{\rotatebox{0}{\includegraphics{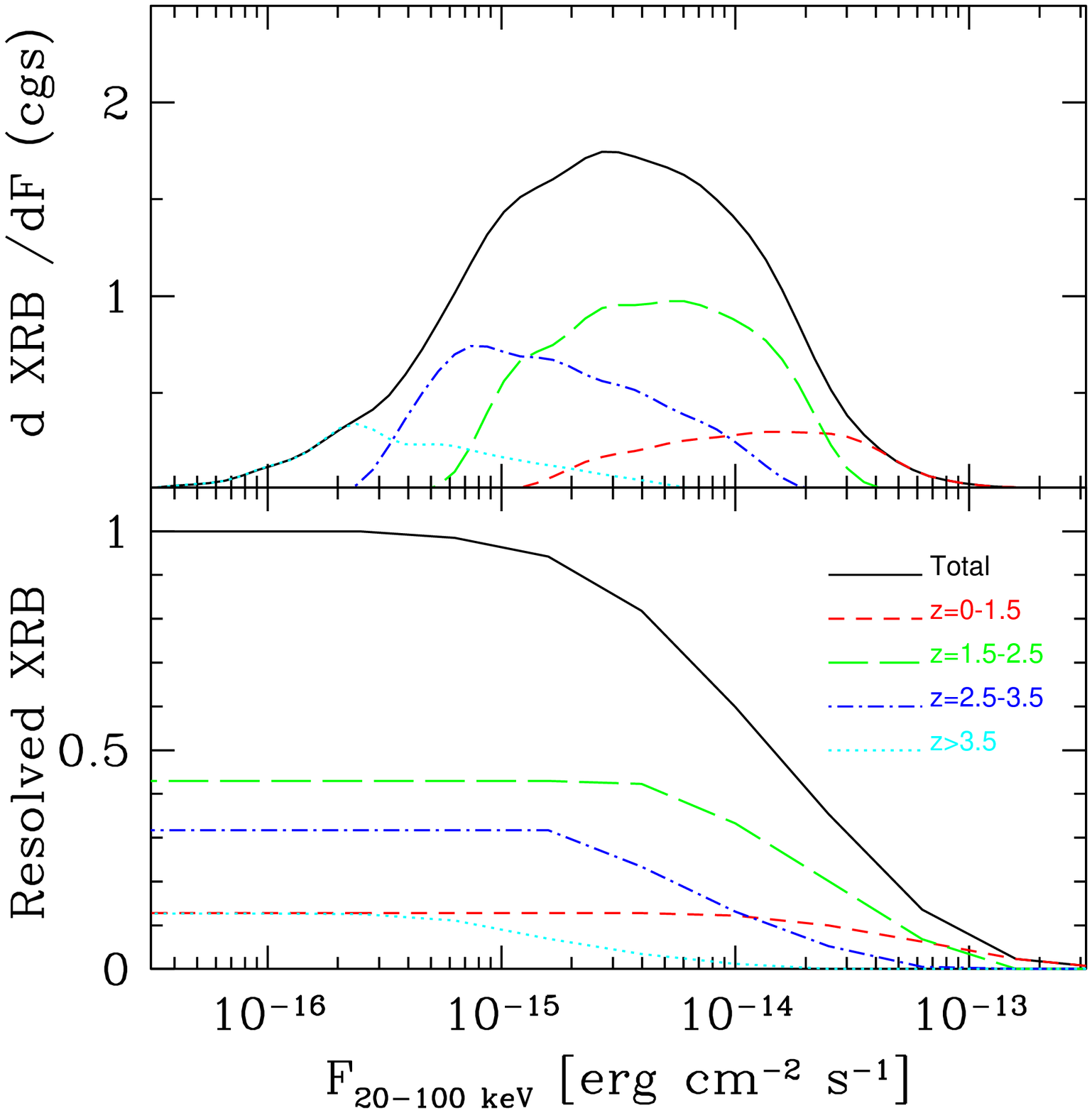}}}
\end{center}
{\footnotesize
\vspace{-0.1cm } Fig. 6. - Top panel. The 20-100 keV flux distribution
of AGNs contributing to the CXB at 30 keV for different bins of
redshift (see labels in the bottom panel). Bottom panel. The fraction
of the CXB contributed by sources with brighter than the value in
abscissa, for the different redshift shells.
\vspace{0.3cm}}

Fig. 7 shows the source counts expected in our model in the 13-80 keV
band, to be compared with the HEAO1-A4 data point (Levine et
al. 1984); .  note that the predicted counts agree very well with the
bright flux point. We also compare our predictions with the
extrapolation to the 13-80 keV band of the BeppoSAX HELLAS 5-10 keV
counts, assuming spectral shapes depending of different values of the
adopted absorbing columns.  This comparison suggests that the bulk of
the hard X-ray selected sources with flux at a few $10^{-13}$ erg
cm$^{-2}$ s$^{-1}$ has a substantial absorbing column, of the order of
$N_H=10^{23}$ cm$^{-2}$.

So far we have compared the predictions of our model with integrated
observations, i.e., the spectrum of the CXB at energies higher than 30
keV and the hard X-ray counts.  We now proceed to compare the model
with observations of differential nature in redshift and in
luminosity, which can provide tighter constraints. Here the comparison
is more critical due to the difficulty of obtaining well defined,
complete samples of highly obscured AGNs at redshifts $z\approx
2.5-3$.  The most recent results on the number and luminosity density
of AGNs come from Chandra, XMM-Newton deep, pencil beam surveys: the
North and South Chandra Deep Fields; Brandt et al. 2002; Barger et
al. 2002; Giacconi et al. 2002; Cowie et al. 2003; the Lockman Hole
survey (Hasinger 2003). We shall compare with the wider but shallower
surveys from XMM-Newton (Fiore et al.  2003). In particular, the
latter authors combined the results of the optical identification of
the HELLAS2XMM 1dF survey with carefully selected identifications from
deep Chandra and XMM surveys to obtain a well defined, flux limited
sample of 317 hard X-ray selected sources (2-10 keV), 70 \% of them
with a measured redshift.  These observations are performed in the
$2-10$ keV band; hence they imply generally small uncertainties
affecting the correction for line of sight obscuration in the
derivation of rest frame luminosity.

\begin{center}
\vspace{0cm} 
\scalebox{0.45}[0.45]{\rotatebox{0}{\includegraphics{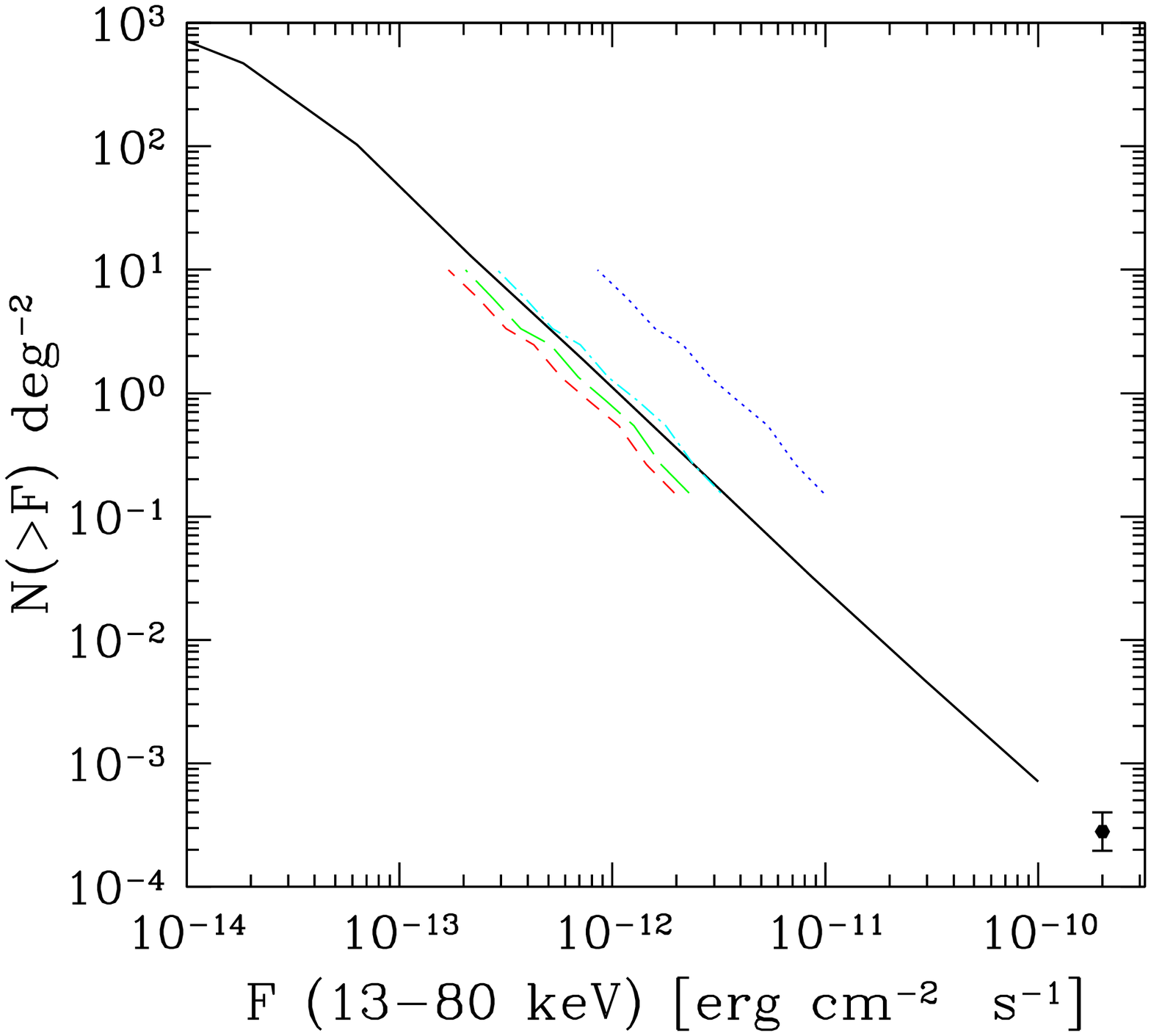}}}
\end{center}
{\footnotesize
\vspace{-0.1cm } Fig. 7. - The source counts of hard X-ray AGNs
contributing to the CXB is shown for the band 13-80 keV, and compared
with the HEAO1-A4 data point (Levine et al. 1984).  We also show the
extrapolation to the 13-80 keV band of the BeppoSAX HELLAS 5-10 keV
counts on assuming different absorbing column densities: $N_H=10^{21}$
cm$^{-2}$ (short-dashed line), $N_H=10^{23}$ cm$^{-2}$ (long-dashed
line), $N_H=10^{23.5}$ cm$^{-2}$ (dot-dashed line), $N_H=10^{24}$
cm$^{-2}$ (dotted line)
\vspace{0.3cm}}

In Fig. 8 we compare our predictions with the evolution of the number
and luminosity densities of AGNs in three luminosity bins, estimated
by Fiore et al.  (2003). Note that the three luminosity bins adopted
by Fiore et al. (2003) for statistical reasons differ from those
adopted in Fig. 2, which were chosen to single out the class of
objects producing the dominant contribution to the CXB.

All the predicted densities drop substantially from $z\approx 2$ to
the present. The agreement with the data is excellent for the highest
luminosity bin, not surprisingly since such objects are also sampled
in the optical band, where the model was already successfully tested in
Menci et al. (2003). In addition, Fiore et al. (2003) find a nice
agreement between their data for high luminosity sources and the
evolution of optically selected AGNs with $M_B<-24$ estimated by
Hartwick \& Shade (1990). This agreement confirms that, at least for
the very luminous AGNs, the bolometric corrections adopted in the B
and in the X-ray band are fully consistent.

At lower luminosities, the decline for $z<1-2$ is less pronounced in
the predictions as well as in the observations. In the former, this is
due to the larger quantity of galactic cold gas left available for
accretion in the less massive galaxies. This is a natural feature in
hierachical scenarios, since more massive potential wells originate
from clumps collapsed earlier in {\it biased} regions of the
primordial perturbation field; the higher densities then prevailing
allowed for earlier condensation and hence enhanced star formation at
high redshifts. Thus, at low $z$ a larger fraction of cold gas will
have already been converted into stars, and both star formation and BH
accretion are considerably suppressed.  We note though that the
decrease of the peak redshift with decreasing luminosity appears to be
significantly smaller than indicated by the data.  In particular,
at z$=1-2$ the observed density of Seyfert-like AGNs is a factor
$\approx 2$ lower than predicted by the model; a similar
difference is present also for the intermediate luminosity objects
($L_{2-10}=10^{44-44.5}$ erg s$^{-1}$) in the redshift bin $z= 2-4$.

\begin{center}
\vspace{0.2cm} 
\scalebox{0.46}[0.46]{\rotatebox{0}{\includegraphics{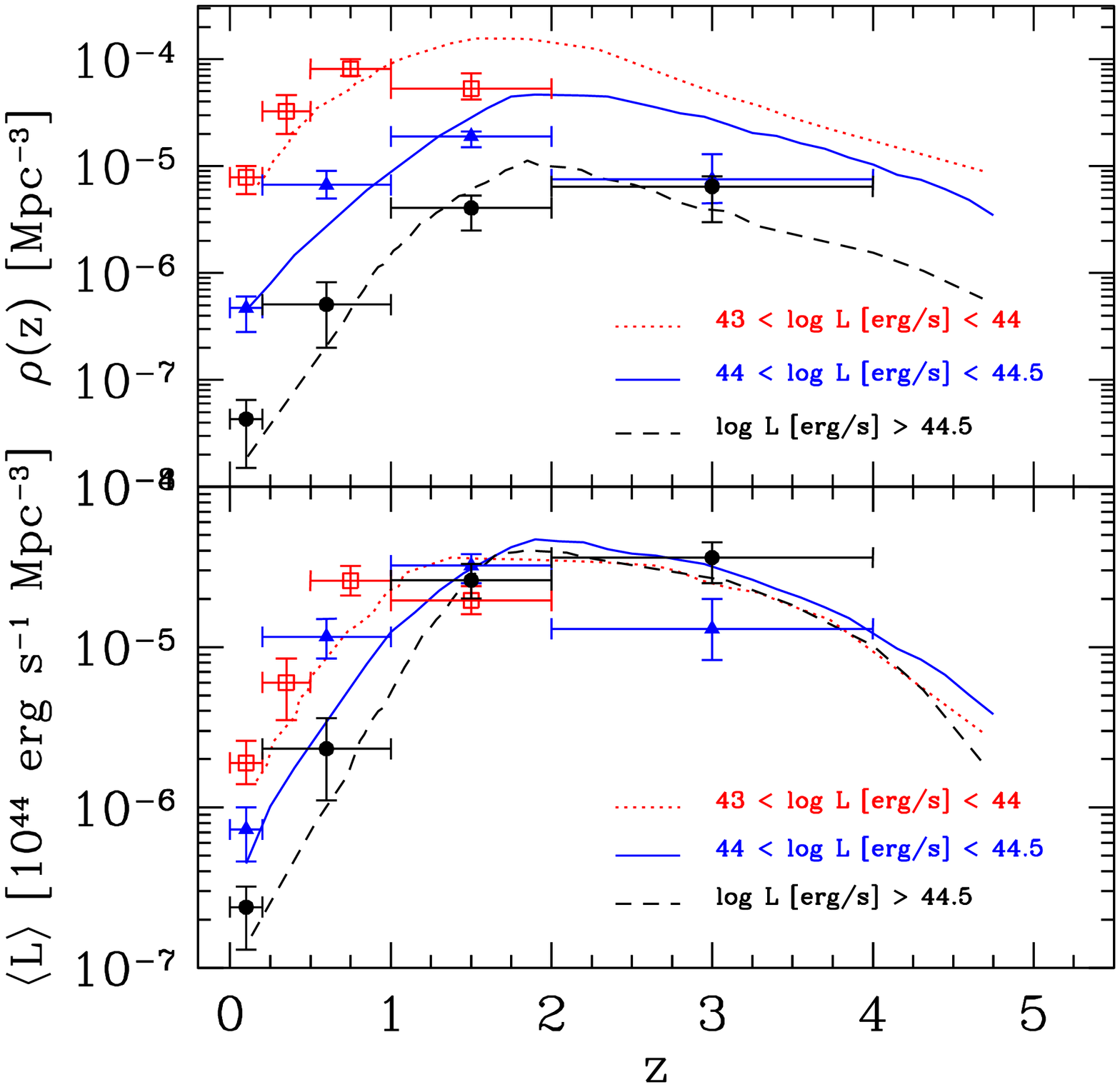}}}
\end{center}
{\footnotesize
\vspace{-0.1cm } 
Fig. 8. - Top: The evolution of the number density of X-ray AGNs 
in three bins of luminosity (in units of erg/s, in the band 2-10 keV): 
$43<log L_X <44$ (dotted line), $44<log L_X<44.5$ (solide line),  
$44.5<log L_X$ (dashed line). The data for the above luminosity bins 
(squares, triangles, and circles, respectively) are taken from Fiore et al. (2003). 
Bottom: The evolution of the X-ray luminosity density 
for the same luminosity bins.  
\vspace{0.2cm}}

The reason for such a discrepancy can be traced back to the shape of
the high-$z$ X-ray luminosity function. 
This is shown in fig. 9, where we compare our model results with  
the observational LFs derived by Fiore et al. (2003, upper panel) and by Ueda et al. 
(2003, lower panel), 
which are obtained from a combination of HEAO1, ASCA and {\it Chandra} data, and 
extend down to lower luminosities. The above observational results 
concur to indicate that the LFs at $z\gtrsim 2$ are appreciably 
flatter than at $z=0.5-1$. When the above data are compared to our results, 
a substantial agreement is found at low $z$, while at $z\gtrsim 1.5-2$ 
the model overestimates the number of low luminosity objects found in both 
the observational analysis. Such a 
mismatch can not be reduced by tuning the bolometric correction 
$c_{2-10}$ adopted in our model, since the latter affects only the 
normalization of the luminosities. 

\begin{center}
\vspace{0cm} 
\scalebox{0.48}[0.48]{\rotatebox{0}{\includegraphics{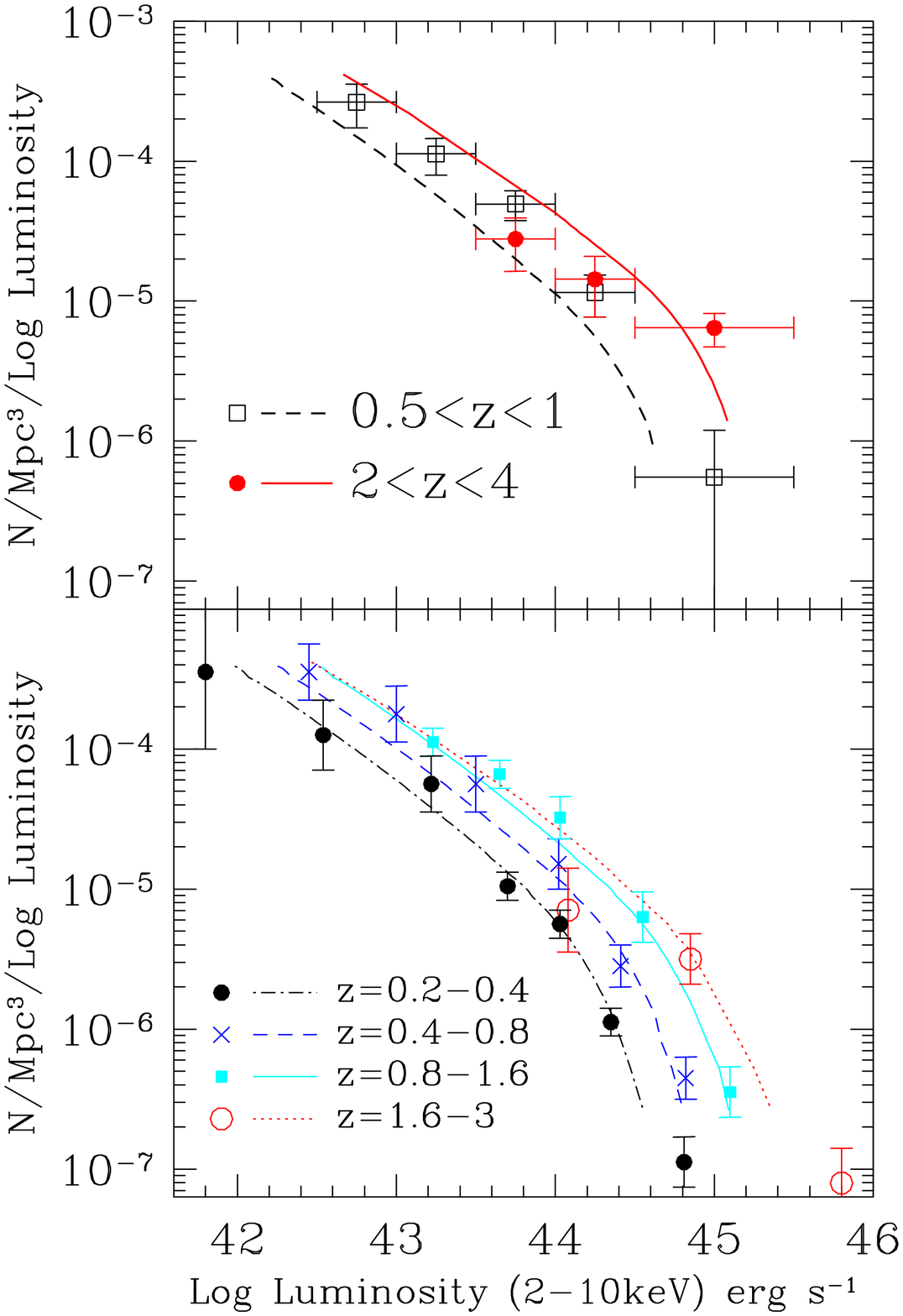}}}
\end{center}
{\footnotesize
\vspace{-0.1cm } Fig. 9. - The predicted LFs in the energy range 2-10
keV at low redshifts $0.5<z<1$ (dashed line) and high redshifts $2<z<4$ (solid line)
are compared with observational values derived from the same sample used in 
Fiore et al. (2003) to derive the densities in Fig. 8. 
\vspace{0.3cm}}

\section{Discussion}

We have incorporated the description of the X-ray properties of AGNs
into the hierarchical picture of galaxy evolution. Our semi-analytic
model, already proven to match the observed evolution of luminous
optically selected QSOs over the redshift range $0<z<6$ (Menci et
al. 2003), is here extended to bolometric luminosities $L$ a factor 10
lower. So we describe the history of accretion down to $L\sim 10^{45}$
erg/s, for which the main observational information comes from the
X-ray band.

We have compared our model with X-ray observations either corrected
for gas obscuration, or performed in the hard ($E>30$ keV) band
not affected by photoelectric absorption.

We find that our model is encouragingly able to match the level of the
cosmic X-ray background (CXB) at 30 keV (Fig. 2). We predict that the
largest contribution (around 2/3) to the CXB comes from intermediate
luminosity sources $43.5<log (L_X/erg\,s^{-1}) <44.5$, and that 50 \%
of its total specific intensity is produced at $z<2$, see figs. 2,
3. The predicted $30-300$ keV CXB spectrum shown in Fig. 3 has been
computed with the simplifying assumption of a fixed spectrum for all
individual souces, and neglecting the absorption in Compton-thick
sources. If a spread in the spectral slope $\alpha$ and in the
high-energy cutoff $E_c$ is included, a harder spectrum would
naturally obtain.  Concurrently, if the distribution of Compton-thick
sources is skewed towards larger redshifts, the absorption would not
affect the spectrum at large energies ($E\gtrsim 100$ keV), where the
CXB is contributed mainly by low- redshifts sources; this results in
an effective flattening of the spectrum.  Since the inclusion of both
these effects in the model would lead to a harder spectrum, a
confirmation of the slope masured by HEAO1-A4 with future experiments
would be of major interest.

We also predict (figs. 6, 7) the number counts and the flux
distribution at different $z$ in the hard (20-100 keV) X-ray band for
the sources contributing to the X- ray background. We have shown that
$\sim 50$ \% of the CXB at 30 keV is produced by sources brighter than
$2\,10^{-14}$ erg cm$^{-2}$ s$^{-1}$ in the 20-100 keV band. Such
predictions can provide guidelines for aimed observations and future
experiments.

When compared to the observed evolution of the number and luminosity
density of AGNs with different $L_X$ (Fig. 8), our model
agrees with the observations concerning all luminosities $L_X>10^{43}$
erg/s for low or intermediate redshifts $z\lesssim 1.5-2$. Thus, adopting
a universal spectrum (and hence fixed bolometric
corrections) our model matches the observations concerning both the
optical (see Menci et al. 2003) and the X-ray band, at least for
objects with bolometric luminosities $L\gtrsim 10^{46}$ erg/s; 
in particular, the density of luminous ($L_X>10^{44.5}$ erg/s) AGNs 
peaks at $z\approx 2$, while for the low luminosity sources
($10^{43}<L_X/{\rm erg\,s^{-1}}<10^{44}$) it has a broader maximum
around $z\approx 1.5$; the decline from the maximum to the value at
the present epoch is around 3 dex for the former class, and 1.5 dex
for the latter class.
At larger redshifts $z\gtrsim 2$, the model still reproduces the observed
number and luminosity densities of AGNs stronger than $10^{44.5}$
erg/s, but at z$=1-2$ the predicted density of
Seyfert-like AGNs is a factor $\approx 2$ larger than observed; 
a similar difference is present also for the intermediate
luminosity objects ($L_{2-10}=10^{44-44.5}$ erg s$^{-1}$) in the
redshift bin $z= 2-4$. We next
discuss our interpretation of both the low-$z$ and the high-$z$
results.

For $z\lesssim 2$, the model results agree with the observed number
and luminosity densities in indicating a drop of the AGN population
for $z<2$ which is faster for the strongest sources. In our picture 
the drop is due to the combined effect of: 1) the decrease of the
galaxy merging and encounter rates which trigger the gas
destabilization and the BH feeding in each galaxy; 2) of the decrease
of the galactic cold gas, which was already converted into stars or
accreted onto the BH. The faster decline which obtains in massive
galaxies (and hence for luminous AGNs) is related in particular
to the latter effect. Indeed, in hierarchical clustering scenarios the
star formation history of larger objects peaks at higher $z$, since
massive objects originate from progenitors collapsed in biased regions
of the Universe where/when the higher densities allowed for earlier
star formation; so, at low $z$ such objects have already exhausted most 
of their gas. On the other hand, less massive galaxies
are continuously enriched by low-mass satellites, whose star formation
is more smoothly distributed in $z$, and which retain even at
$z\approx 0$ an appreciable fraction of cold gas available for BH
accretion.

For $z\gtrsim 2$, the slight excess of the number and luminosity
densities of weak AGNs with $L_X<10^{44.5}$ erg/s over the
observations may originate both in the observations and in the
modelling.

On the one hand, data incompleteness is to be expected at high $z$ for
low- luminosity objects. Althouh the data are corrected for
absorption, the sources would be lost from the sample when heavily
obscured below the detection limit. Also, the observed number and
luminosity densities do not include Compton-thick sources whose
contribution is instead included in the model predictions. If the
amount of obscuring gas increased with redshift, as would be the case
if a large amount of gas accumulates in the surroundings of Eddington
limited BHs, lower-$z$ and higher luminosity sources would be effectively
favoured.  Future observations extended to harder bands will clarify
the issue; in the meanwhile, we plan to improve the comparison between 
the model and the data through the inclusion in the model of gas absorption 
and the computation of the related effect of Compton thick sources. 
This in principle can be done self-consistently, since the amount of 
galactic gas is predicted in our model (see fig. 1 in Menci et al. 2003). 

On the other hand, the excess of predicted low-luminosity AGNs at high $z$ could 
originate from an incomplete modelling in the low mass regime at high $z$. For 
example, the mismatch may be lifted by a mean AGN lifetime increasing with mass 
(see Yu \& Tremaine 2002), a feature that can be implemented in the developments 
of our present fiducial model, specifically through a mass 
dependence of $\tau$ in eqs. (3) and (4). Additional improvements may concern 
the following sectors:

i) The properties of the accretion
disks at low accretion rates. Advection-dominated accretion flows
might constitute an example of such a physics, though they ought to
decrease the emitted luminosity preferentially at high $z$ to explain
the excess. 

ii) The regulation governing the amount of cool gas in
low-mass host galaxies, like the Supernovae feedback. This still
represents the most uncertain sector of all SAMs; it is known to play
a key role in flattening the faint end of the optical galaxy
luminosity functions, since large feedback would deplete the cold gas
reservoirs preferentially in shallow potential wells. 

iii) The 
statistics of DM haloes. The most recent N-body simulations point
toward a mass function of galactic DM clumps somewhat flatter at the 
small-mass end than previously
assumed (see Sheth \& Tormen 1999, Jenkins et al. 2001). 
While this could contribute to solve the issue, the flattening 
is not large enough to explain out the observed excess. 

All the above points can concurr in determining the excess of the predicted 
low-luminosity AGNs at high $z$. We note, however,   that overprediction of the 
low-luminosity sources is a long-standing problem which also affects the number 
and luminosity distribution of faint, high-$z$ galaxies. This would point toward 
an origin of the mismatch in the physics of galaxy formation rather than in the 
description of the accretion processes. If this is indeed the case, tuning the 
free parameters in the SAM (like, e.g., the efficiency of the Supernovae 
feedback, the one which mainly suppress the AGN and star formation activity in 
low mass systems) does not constitute a valid solution for the excess of low-
luminosity X-ray AGNs at high $z$; in fact, the required adjustments would 
worsen the matching between the model results and the observed properties of 
galaxies, like, e.g., the Tully-Fisher relation. On the other hand, implementing 
an {\it ad hoc} parametrical dependence on $z$ of the Supernovae feedback could 
significantly reduce the excess; but in the absence of a physical motivation 
this would not lead to a deeper insight into the galaxy-AGN connection. 

Thus, a real improvement in the modeling requires the inclusion of additional 
physical processes in the SAM (and in particular in the sector concerning the 
feedback) rather than the tuning of the parameters in the existing framework. 
One such process could be well constituted by  the inclusion into SAMs of the 
feedback produced by the AGNs emission itself. Since the AGN activity strongly 
increases with redshift, this could significantly contribute to expell/reheat 
part of the galactic cold gas reservoir in low-mass systems at high $z$.  While 
the modeling of such impulsive processes is particulary delicate, some steps in 
this direction have already been taken (see, e.g., Haenhelt, Natarajan \& Rees 
1998; Silk \& Rees 1998; Wyithe \& Loeb 2003; Cavaliere, Lapi \& Menci 2002; and 
references therein). We shall investigate the effects of such processes 
on the evolution of the AGN population in a next paper. 

In sum, despite the uncertain origin of the overprediction of faint X-ray 
sources at $z\gtrsim 2$, the present model provides a good baseline to include 
the evolution of galaxies and AGNs in the same global picture, being supported 
by a remarkable agreement with the observations of its predictions for brighter 
sources in a wide range of redshifts (from $0<z<6$) and of wavelengths (from 
optical to X-rays). The most distinctive feature of such a picture is the 
dramatic decrease of the AGNs luminosities at $z\lesssim 2$ especially in 
massive galaxies (see Fig. 1), naturally resulting from to the exhaustion of 
cold gas necessary for feeding both the accretion and the star formation;  
relatedly, massive galaxies are predicted to undergo a nearly passive evolution 
from $z\approx 2$ to the present. The relevance of such an exhaustion in determining the
observed properties of the AGN population (in both the optical and the X-rays) is confirmed 
by recent N-body simulations (Di Matteo et al. 2003). 
The above picture thus naturally explains the parallel evolution of BH accretion 
and star formation in spheroidal systems; this, originally discussed by Monaco, 
Salucci \& Danese (2000) and Granato et al. (2001), is supported by recent 
works (see Franceschini, Hasinger, Miyaji, Malquori 1999; 
Haiman, Ciotti \& Ostriker 2003) which also enlightened its simultaneous 
consistence with the evolution of the optical and the X-ray luminosity functions 
of AGNs (Cattaneo \& Bernardi 2003). 
The physical origin of such a parallel evolution is here clarified and shown to arise as a 
natural outcome of hierarchical galaxy formation.

\end{document}